\documentclass{pas}

\usepackage{multirow}
\usepackage{amssymb}
\usepackage{subcaption}

\begin{document}

\lefttitle{Publications of the Astronomical Society of Australia}
\righttitle{Southon et al.}

\jnlPage{1}{4}
\jnlDoiYr{2025}
\doival{10.1017/pasa.xxxx.xx}

\articletitt{Research Paper}

\title{Radial velocities and stellar populations for a sample of MATLAS survey dwarfs}

\author{\gn{Kate} \sn{Southon}$^{1, 2}$, \gn{Jonah S.} \gn{Gannon}$^{2}$, \gn{Duncan A.} \sn{Forbes}$^{2}$, \gn{Arsen} \sn{Levitskiy}$^{2}$, \gn{Maria Luisa} \sn{Buzzo}$^{2}$, \gn{Aaron J.} \sn{Romanowsky}$^{3, 4}$ and \gn{Jean P.} \sn{Brodie}$^{2, 4}$}

\affil{$^1$ School of Physical and Chemical Sciences – Te Kura Mat\={u}, University of Canterbury, Private Bag 4800, Christchurch 8140, New Zealand \\
$^2$ Centre for Astrophysics and Supercomputing, Swinburne University, John Street, Hawthorn VIC 3122, Australia\\
$^{3}$ Department of Physics and Astronomy, San Jos\'e State University, One Washington Square, San Jose, CA 95192, USA \\
$^{4}$ Department of Astronomy \& Astrophysics, University of California Santa Cruz, 1156 High Street, Santa Cruz, CA 95064, USA}

\corresp{K. Southon, Email: kasouthon@gmail.com}

\citeauth{Southon et al., 
Radial velocities and stellar populations for a sample of MATLAS survey dwarfs, {\it Publications of the Astronomical Society of Australia} {\bf 00}, 1--12. https://doi.org/10.1017/pasa.xxxx.xx}

\history{(Received xx xx xxxx; revised xx xx xxxx; accepted xx xx xxxx)}

\begin{abstract}
Spectroscopic observations are essential for confirming associations, measuring kinematics, and determining stellar populations in dwarf galaxies. Here, we present Keck Cosmic Web Imager (KCWI) spectra for 12 MATLAS survey dwarfs. For 9, we confirm recession velocities consistent with their literature-assumed host galaxies. 
We propose revisions of the host galaxy associations for MATLAS-631, 1494, and 1938. For MATLAS-1494, our measured redshift reclassifies it from an ultra-diffuse galaxy candidate to a dwarf galaxy that is of smaller physical size and places it in the field. It also appears old and passive, providing a challenge to models that invoke quenching by tidal effects. Additionally, we measure stellar population estimates for 7 of the 12 galaxies, finding 
a `mixed bag' of old quenched galaxies and those that are currently forming stars. Compared to the literature we find generally younger ages and higher metallicities. This result may help reconcile the observed offset of MATLAS survey dwarf galaxies from the universal stellar mass–metallicity relationship reported by Heesters et al. (2023). 
\end{abstract}

\begin{keywords}
galaxies: distances and redshifts, galaxies: dwarf, galaxies: evolution, galaxies: general
\end{keywords}

\maketitle

\section{Introduction}

Ultra diffuse galaxies (UDGs) are a subclass of low surface brightness galaxies that are defined by a distance-independent central surface brightness limit ($\mu_{g,0}$ $\ge$ 24 mag per sq. arcsec) and a distance-dependent effective radius ($R_\mathrm{e}$ $\ge$ 1.5 kpc; \citealp{vandokkum2015}). In addition to their large size for their low mass (stellar masses are typically 10$^7$ to 10$^9$ M$_{\odot}$   \citep{DiCintio2017}), they can host many more globular clusters (GCs) than an equivalent classical dwarf galaxy of the same stellar mass \citep{Forbes2020, Saifollahi2022, Marleau2024}. 

UDGs are found in all environments, with their abundance scaling roughly linearly with the group/cluster halo mass \citep{Janssens2019, LaMarca2022}. 
Going from low to high density, they become, on average, older, redder, gas-free and host more globular clusters \citep{Prole2019, FerreMateu2023, Buzzo2024, Buzzo2025}. For example, the mass contained in the GC system relative to the stellar mass of the galaxy (M$_{GC}$/M$_{\ast}$) increases from 0.0\% to 1.2\% to 3.4\% for low, intermediate and high density environments respectively \citep{Buzzo2025}. 

Redshifts are clearly important in determining the environment of a galaxy. When redshifts are not available we can use statistical methods to associate a sample of UDGs with a given cluster, e.g. requiring that they are consistent with the low mass end of the cluster red sequence \citep{LaMarca2022} or that their surface density falls off with distance from the cluster centre \citep{Janssens2019}. For UDGs in small groups or the field, such methods are difficult to apply. In this case, UDGs are often just assumed to be at the redshift of the group \citep{Poulain2021}
or even simply assigned a distance such that $R_\mathrm{e}$ $\ge$ 1.5 kpc, i.e. the minimum size to meet the UDG definition of
\citet{Zaritsky2023}. Thus, although spectroscopic redshifts for individual faint UDGs are time consuming, they are also important to confirm their UDG status and place them in the correct environment. The catalog of \citet{Gannon2024} and updates summarises the spectroscopic studies of UDGs' stellar populations and internal dynamics, as drawn from the literature. These galaxies all meet the UDG definition and have either a velocity dispersion or stellar population parameters spectroscopically measured. According to the studies of 
\citet{2017ApJ...838L..21K} and 
\cite{FerreMateu2023} UDGs in clusters tend to be old (mean age $\sim$ 8.3 Gyr) and metal-poor ([M/H] $\sim$ --1.0 dex). This work should be complemented by the study of additional field and group UDGs with stellar population measurements. 

Here we present new spectroscopic redshifts using Keck/KCWI for UDG candidates and dwarf galaxies from the MATLAS survey \citep{Poulain2021, Marleau2021}. The MATLAS survey imaged low-to-intermediate density environments using the Canada-France-Hawaii Telescope (CFHT), finding 59 UDGs and a large number of classical dwarf galaxies \citep{Habas2020}. Subsequent follow-up included HST imaging of their GC systems \citep{Marleau2024} and a VLT/MUSE campaign to test the association of the galaxies with their assumed hosts and to probe their stellar populations \citep{Heesters2023}. 
The latter gave redshifts for 56 UDGs and classical dwarf galaxies, confirming the association of 75\% (79\% for dwarf ellipticals) of them with their assumed hosts. The galaxies were found to be generally very old ($\sim$10 Gyr) and metal-poor ($\sim$ --1.5 dex). No galaxies were found to have average ages of less than 6 Gyr, including those with a recent burst of star formation as revealed by emission lines. Compared to Local Group dwarfs \citep{Simon2019}, their galaxies have systematically lower metallicities for their stellar mass. 
Stellar populations of MATLAS UDGs were also studied in \citet{Buzzo2024}. In their spectral energy distribution fitting process, they fixed the redshift to that of the assumed host galaxy velocity in the cases where the redshift of the galaxy itself was unknown. Their sample was found to have on average younger ages and higher metallicities compared to the spectroscopic \cite{Heesters2023} study.
They also presented Keck/DEIMOS spectra for three UDGs (MATLAS-342, 368, 1059) finding all three were consistent with their host galaxy, and bona fide UDGs. The subsequent work of \cite{Buzzo2025} extended the SED fitting to 
dwarfs that were slightly brighter and/or slightly smaller in half-light radius than UDGs, referred to as `nearly-UDGs' (NUDGes). They found that NUDGes are similar to UDGs in all properties except for size and brightness, with a difference in GC content supporting a division into two formation classes. We also note the work of \cite{Rong2020}, who stacked SDSS spectra of 28 UDGs in low density environments i.e. residing outside of the virial radius of the nearest galaxy cluster. Their UDGs were selected with $\mu_{g,0} > 23.5$ mag arcsec$^{-2}$, $R_\mathrm{e} > 1.5$ kpc, and $\log(M_*/M_\odot) > 9.0$. They measured a mean mass-weighted age of 5.2 $\pm$ 0.5 Gyr and [M/H] = --0.82 $\pm$ 0.14 dex.

In this study, we aim to probe the redshifts and stellar populations of more MATLAS dwarfs both to compare with the literature for the sources with measurements available, as well as to add information about the galaxies that have not previously been studied.
In Section 2 we describe the data acquisition, reduction and basic analysis of our KCWI spectra. 
Section 3 gives the results and discussion. Conclusions are given in Section 4. We assume a Hubble constant of 72 km/s/Mpc.

\section{KCWI Data}
\subsection{Data Acquisition and Reduction}

\begin{figure*}
    \centering
    \includegraphics[width =\textwidth]{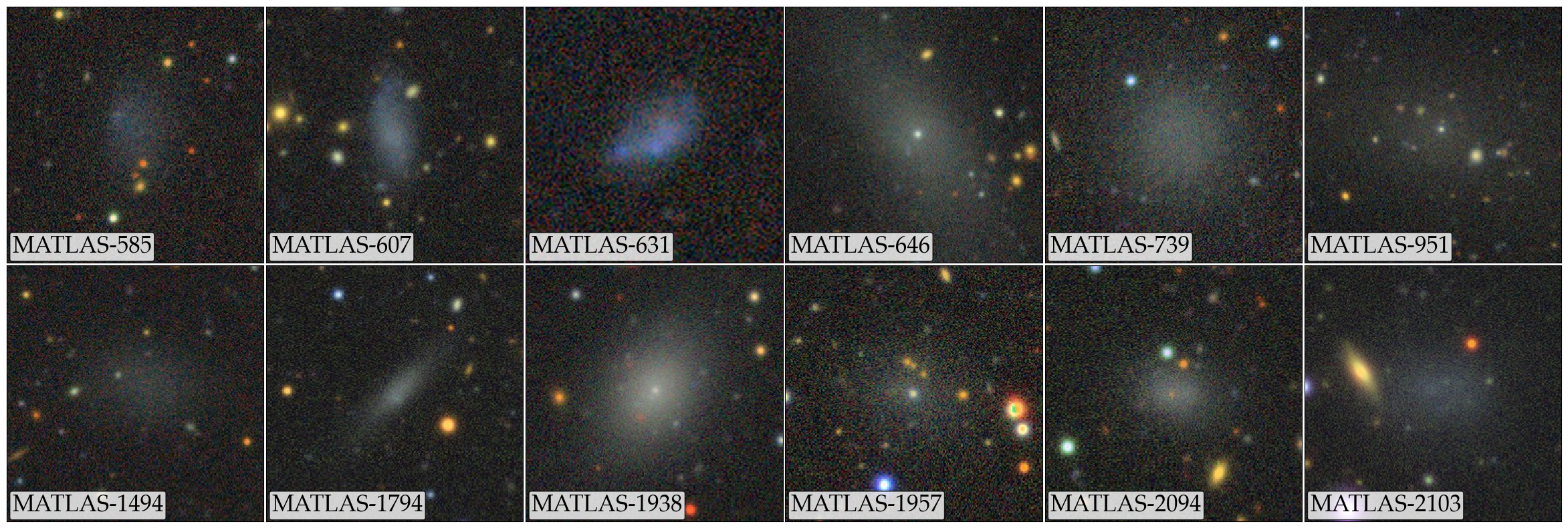}
    \caption{Processed postage stamps of the MATLAS galaxies analysed in this study. Each stamp is 1 arcmin on each side. In all panels, north is up and east is left. The RGB images are constructed using the $g$, $r$ and $i$ bands from DECaLS. MATLAS-585, 607, 631 and 2103 are blue and show hints of ongoing star formation.}
\label{fig:stamps}
\end{figure*}

The majority of data were observed on the night of 2024, February 13th. Conditions were clear with 1'' seeing. On this night the Keck II telescope suffered a fault with its lower shutter and dome rotation control limiting observations to targets above an elevation of 39 degrees and within a small sliver ($\sim \pm 15$ deg) of an azimuth of 130 degrees. With these limitations, we pursued a program of observing galaxies from the MATLAS survey \citep{Habas2020} using the Keck Cosmic Web Imager (KCWI). The galaxies are a mix of UDG candidates and classical dwarfs (many with a central surface brightness brighter than the UDG limit), that matched the restricted observing constraints. Images of the sample galaxies are shown in Figure \ref{fig:stamps}.

KCWI was configured using the BL grating with a central wavelength of 4550 \AA~ and the medium slicer. In this configuration, the resolution is R = 1800, and the field of view is 16.5'' $\times$ 20.4''. Observations were taken simultaneously in the red arm, but not used in this analysis.
One galaxy of the sample was observed on the night of 2024 April 14th (MATLAS-1938), in the same Medium/BL/4550 configuration as those observed on 2024, February 13th. 
The observations for each galaxy are summarised in Table \ref{tab:observations}. The Table also includes the assumed host galaxy from \cite{Poulain2021}. 

The data were reduced using the standard KCWI pipeline \citep{Morrissey2018}, disabling sky subtraction. The data were then cropped to a common wavelength range and stacked using the mosaic routine \texttt{Montage} \citep{Montage}. For further details see \citet{Gannon2024b}.

\begin{table*}
\begin{center}
\caption{Summary of observations and KCWI set-up.}\label{tab:observations}
{\tablefont\begin{tabular}{@{\extracolsep{\fill}}lccrrrrrrrr}
\toprule
Target ID & Type & Assumed Host & RA (J2000) & Dec (J2000) &  Date & Exposure [s] & $N_\mathrm{exposures}$\\ 

MATLAS-585 & UDG& IC~0560 & 09:45:49.11 &	-00:32:51.9	 & 14/02/2024  & 300 & 7 \\

MATLAS-607 & dwarf & PGC028887 & 09:59:36.18 & +11:46:29.4 & 14/02/2024 & 300 & 3 \\

MATLAS-631 & dwarf & PGC029321 & 10:06:00.19 & +13:03:31.8& 14/02/2024  & 300 & 2 \\

MATLAS-646 & UDG$^{\ast}$ & NGC~3156 & 10:13:29.62 & +03:26:55.2	 & 14/02/2024  & 300 & 3 \\

MATLAS-739  & dwarf & NGC~3377 &  10:46:24.62 & +14:01:28.6 & 14/02/2024 & 240 & 2 \\

MATLAS-951  & UDG & NGC~3640 & 11:20:31.83 &	+03:13:59.6 & 14/02/2024  & 240 & 5 \\

MATLAS-1494 &  UDG$^{\dagger}$ & NGC~4690 & 12:47:23.24 & -01:24:16.2 & 14/02/2024  & 240 & 3  \\

MATLAS-1794 & UDG & NGC~5507 & 14:13:52.63 & -03:03:21.1 & 14/02/2024  & 240 & 11 \\

MATLAS-1938 & dwarf & NGC~5813 & 15:00:33.02 & +02:13:50.8 & 14/04/2024  & 1200 & 2 \\

MATLAS-1957 & UDG & NGC~5813 &  15:01:54.71 & +01:37:33.0 & 14/02/2024  & 240 & 7 \\

MATLAS-2094 & dwarf & NGC~6010 & 15:53:14.76 & +00:42:35.0 & 14/02/2024  & 240 & 8 \\

MATLAS-2103 & UDG & NGC~6017 & 15:56:20.09 & +06:11:17.6 & 14/02/2024  & 240 & 7
\botrule
\end{tabular}}
\end{center}

\begin{tabnote}
Notes: MATLAS galaxy ID (column 1), galaxy type (column 2) if known UDG, otherwise dwarf galaxy from \cite{Poulain2021} and \cite{Marleau2021}, $^{\ast}$ UDG based on parameters from \cite{Zaritsky2023}, 
$^{\dagger}$ no longer UDG with new distance (see text for details), expected host galaxy (column 3) from \cite{Poulain2021}, MATLAS galaxy coordinates (columns 4,5), Date of KCWI observation (column 6), time of single exposure (column 7), and number of exposures (column 8). 
\end{tabnote}

\end{table*}

\subsection{Data Analysis}
The reduced KCWI data cubes were displayed using QFitsView \citep{QFitsView}.  
In QFitsView, a source region was selected to encompass each galaxy, and a corresponding sky background region was defined from the surrounding area. Bright contaminating sources were manually excluded from the selected regions. Images showing the identified source and background regions for each galaxy are provided in Appendix \ref{appendix:galaxymasks}. On-sky subtraction was performed by subtracting the defined background from the source region, and a 1D spectrum of the galaxy was extracted. 

To measure the recessional velocities and stellar populations of each galaxy, we used the spectrum-fitting code pPXF \citep{Cappellari2017}. Initial redshift estimates were obtained through visual inspection of the prominent absorption features within each spectrum (e.g. Ca H+K, H$\beta$), and subsequently used by pPXF as input for the fitting process. We follow the method described in the pPXF example \texttt{ppxf\_example\_population\_bootstrap.ipynb} \footnote{\url{https://github.com/micappe/ppxf\_examples/blob/main/ppxf\_example\_population\_bootstrap.ipynb}}. When fitting, we use the MILES synthetic stellar library with BaSTI isochrones as templates and include both a 15th order multiplicative polynomial and a first-order additive. The spectra were restricted to the range 3700--5510 Å when fitted to avoid regions contaminated by sky residuals. After an initial fit, 10000 bootstrap fits were performed and final values are quoted as the median of these results with 1-$\sigma$ uncertainties as the 16th and 84th percentiles of the distribution. 
While we include gas templates in the fit process in the cases where emission features are present in the spectrum, results are quoted from the absorption line template fits. The output recession velocities have been barycentric corrected. For MATLAS-585, 607 and 631 we also obtained gas velocities
and found them to be consistent with the absorption line velocity within joint uncertainties. 

\section{Results and Discussion}

\begin{figure}
\includegraphics[scale=.55]{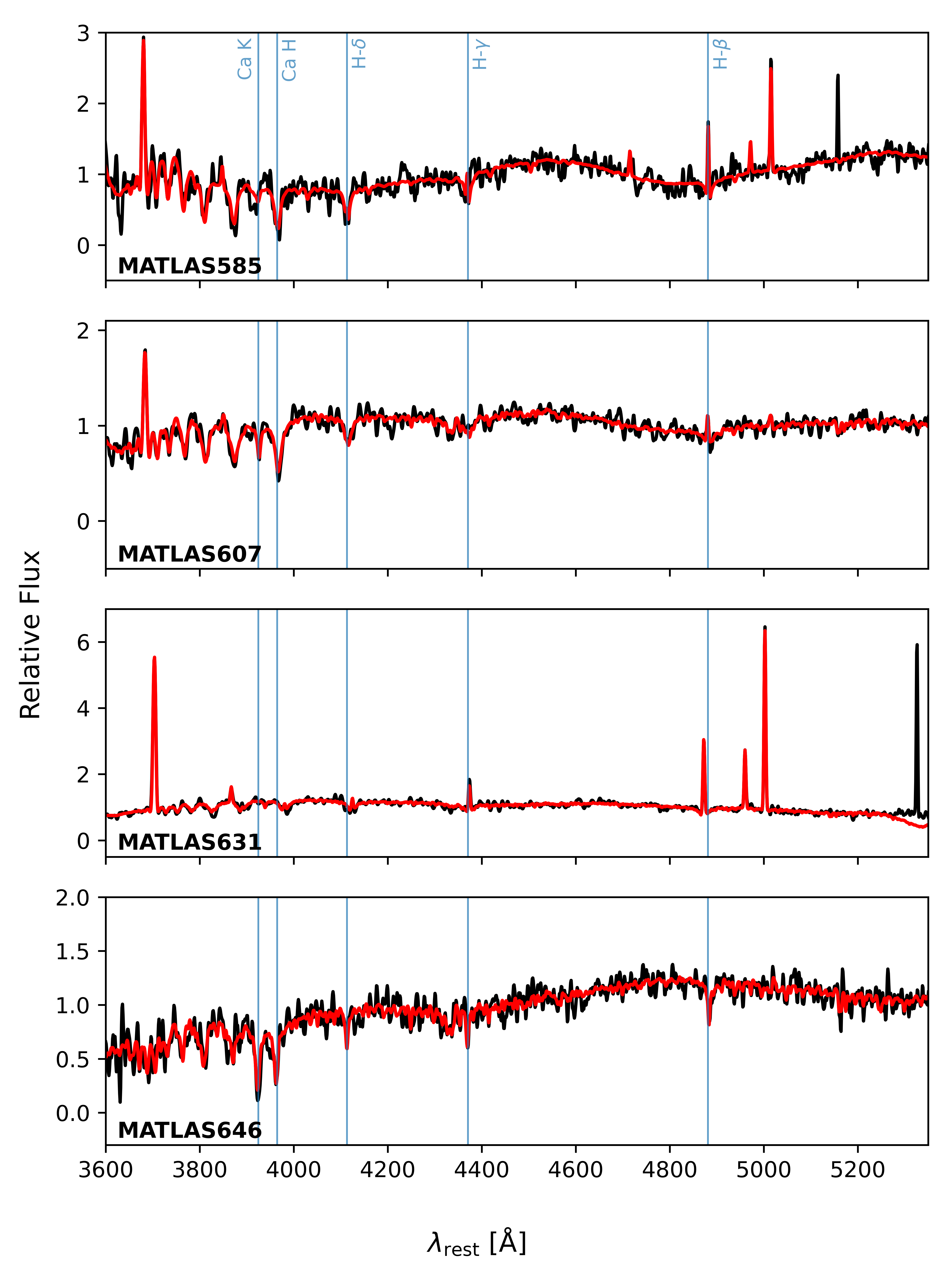}
\caption{KCWI rest frame spectra of MATLAS-585, 607, 631 and 646. The observed spectrum is shown in black, with the pPXF fit overlaid in red. Key absorption lines are shown in blue. MATLAS-585, 607 and 631 also show emission lines (e.g. [OII]3727 and [OII]4959,5007).}
\label{fig:spectra-585-646}
\end{figure}

\begin{figure}
\includegraphics[scale=.55]{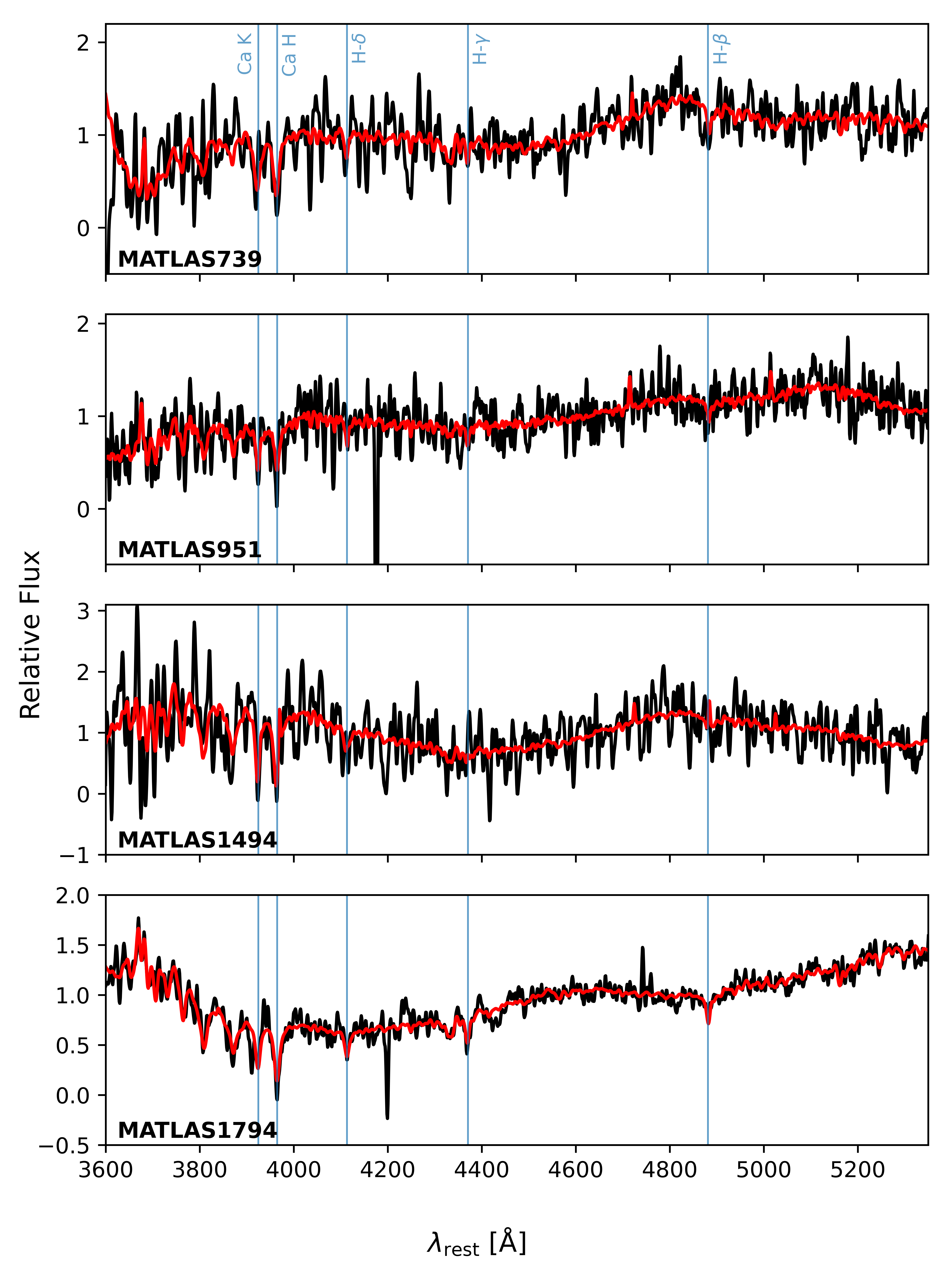}
\caption{KCWI rest frame spectra of MATLAS-739, 951, 1494, and 1794. The observed spectrum is shown in black, with the pPXF fit overlaid in red. Key absorption lines are shown in blue.}
\label{fig:spectra-739-1794}
\end{figure}

\begin{figure}
\includegraphics[scale=.6]{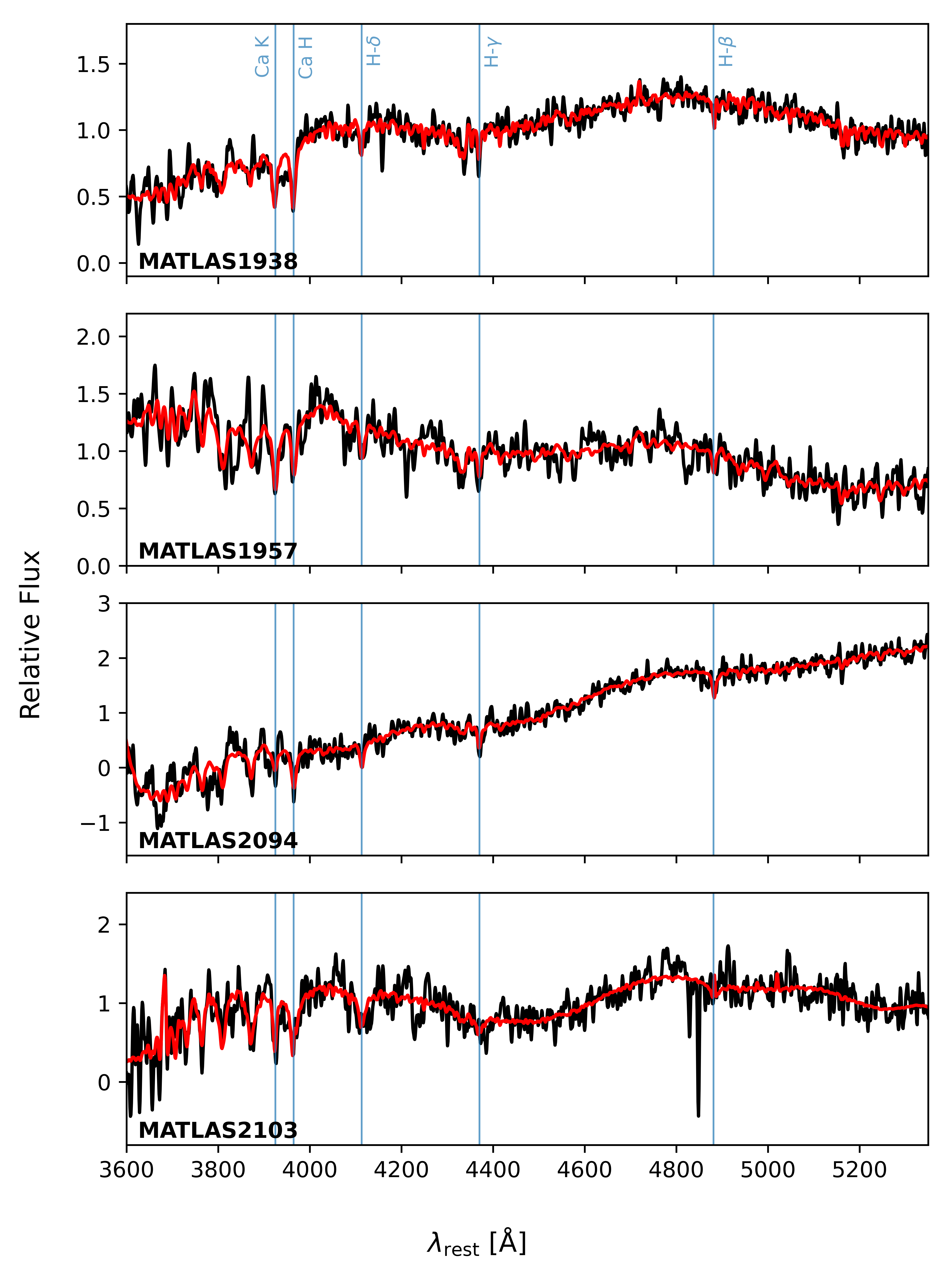}
\caption{KCWI rest frame spectra of MATLAS-1938, 1957, 2094 and 2103. The observed spectrum is shown in black, with the pPXF fit overlaid in red. Key absorption lines are shown in blue. MATLAS-2103 reveals weak emission lines (e.g. [OII]3727).}
\label{fig:spectra-1957-2103}
\end{figure}

The extracted 1D spectra for our MATLAS sample are displayed in Figures  \ref{fig:spectra-585-646}, \ref{fig:spectra-739-1794} and \ref{fig:spectra-1957-2103}. Several galaxies reveal prominent Calcium H+K lines and Balmer series lines in absorption from H$\beta$ to H$\zeta$ and beyond. The Balmer series lines are reminiscent of A stars (i.e. ages of $\sim$ 1 Gyr for solar-like metallicities) but also found in old, metal-poor populations 
(e.g. \citealt{2017ApJ...838L..21K}).
In the some cases emission lines (e.g. [OII]3727, H$\beta$ and [OIII]4959, 5007) are clearly seen, suggesting a young starburst or low ionisation shocks, i.e. MATLAS-585, 607 and 631. Weak features near [OII]3727 may indicate emission in MATLAS-2103, though this cannot be confidently confirmed.
We note that the larger spectroscopic study of \cite{Heesters2023} found 17 of their 56 MATLAS galaxies (30\%) to have emission lines in their MUSE spectrum (which has redder wavelength coverage than our KCWI data). 

\begin{figure}
\includegraphics[width = 0.5\textwidth]{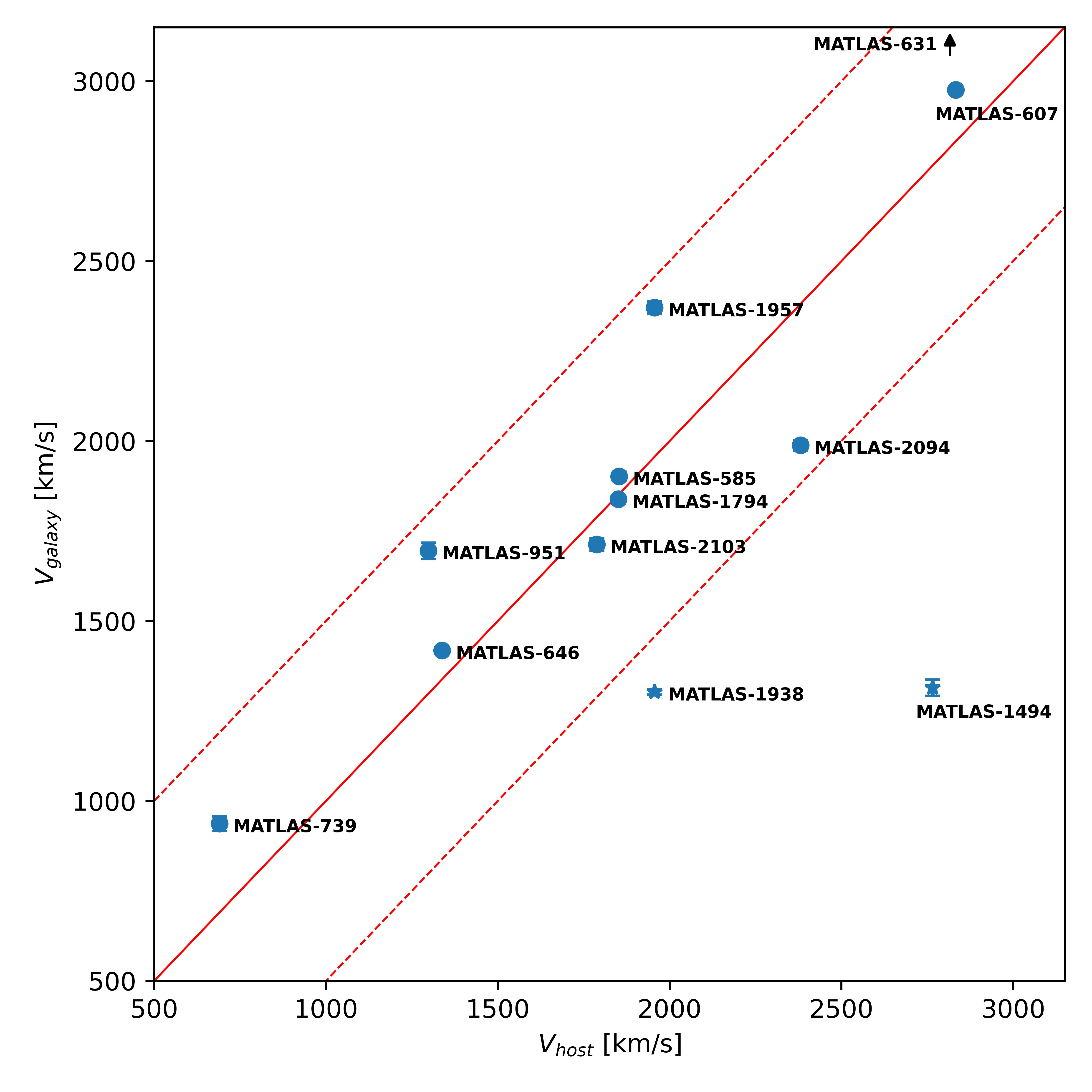}
\caption{Comparison of the fitted recession velocities ($V_{galaxy}$), newly measured from absorption lines in this work, for the MATLAS target galaxies against the literature velocities ($V_{host}$) of their assumed host galaxies. Here, `assumed host galaxy' refers to the larger galaxy the dwarf is thought to be associated with based on projected proximity, as listed in \citet{Poulain2021}. Error bars represent the uncertainties in $V_{galaxy}$. The solid red line corresponds to a perfect agreement between the two velocity estimates, and the dashed red lines represent $\pm$500 km/s from unity. In this work we have revised the host galaxy for MATLAS-1938 and 1494 (here we show their original assumed host with a star symbol). The plot does not show MATLAS-631 which has a recession velocity of $>$10,000 km/s compared to the possible host galaxy PGC029321 with V$_{host}$ = 2816 km/s. Overall we find 10/13 galaxies associated with the host galaxy listed in \cite{Poulain2021}.
}
\label{fig:V-comparison}
\end{figure}

The recession velocity that we measure for each galaxy is plotted against that of the assumed host galaxy \citep{Poulain2021} in Figure \ref{fig:V-comparison}. Informed by the study of 54 small groups by  \cite{2004MNRAS.350.1511O}, who found that group velocity dispersions vary from 25 to 650 km/s with a mean of 260 km/s, we apply a velocity threshold to be associated with a host. In Figure \ref{fig:V-comparison} we show $\pm$ 500 km/s around a line of unity which defines the threshold used by us for the galaxies to be associated with their proposed hosts. By this criterion, we find all galaxies except MATLAS-631, 1494 and 1938 to be associated with the host galaxy suggested by \cite{Poulain2021}. Table 2 lists our recession velocities and we suggest new hosts for MATLAS-631 and 1938, whereas MATLAS-1494 appears to be an isolated field galaxy with no obvious host.  

While all of our spectra are of sufficient quality to obtain a velocity, with typical uncertainty of a few tens of km/s, not all are suitable to derive accurate stellar population parameters. We are unable to derive ages and metallicities for MATLAS-739, 951, 1494, 1794, and 2103. For the remaining galaxies, our pPXF fitting indicates a range of stellar population properties. We find some have ongoing star formation today while for others the main epoch of star formation ceased long ago. The stellar metallicities range from very metal-poor ($\sim$ --2 dex) to almost solar. These mean ages and total metallicities are listed in Table 2. 
For the three strong emission line galaxies, we list their light-weighted stellar populations, given the large and uncertain light-to-mass transformations required. For the other galaxies we give their mass-weighted values. Corner plots showing the distribution of ages and metallicities obtained from our fitting process are in Appendix \ref{appendix:cornerfigs}. 

Table 2 also lists some key galaxy properties such as the number of GCs and the stellar mass of the galaxy. The former are taken from the HST imaging of \cite{Marleau2024} and the latter from the SED study of \cite{Buzzo2024}. While the typical stellar mass has a small range around 10$^8$ M$_{\odot}$, the GC systems vary from very rich (e.g. N$_{GC}$ $\sim$ 29 for MATLAS-1938) to being consistent with no GCs (e.g. MATLAS-1794).

\subsection{Individual Galaxies}

Out of the sample of 12 galaxies, nine have photometry from MATLAS \citep{Poulain2021},
which we use to classify them as dwarfs or UDGs (see definition in Section~1). For the remaining three, we make use of imaging from the Legacy survey, DR10
\citep{Dey2019}.
In the two cases where the automated ``tractor'' galaxy model visually matches the image well (MATLAS-607 and MATLAS-631), we use the corresponding photometric parameters to classify the galaxy (which we have also validated for cases with MATLAS photometry).
The third galaxy, MATLAS-646, does not have a reliable tractor model but does have photometry from the SMUDGes survey \citep{Zaritsky2023} which we use for its classification.
Individual galaxies are discussed in more detail below.

\subsubsection{MATLAS-585, 2094 and 2103}

We have three galaxies in common with \cite{Heesters2023}. They are MATLAS-585, 2094 and 2103. In all three cases, our velocities are in reasonable agreement. It is important to note that the spectral coverage of MUSE is generally redder than that of our KCWI configuration. \cite{Heesters2023} found both MATLAS-585 and 2094 to have emission lines, and MATLAS-2103 to be emission-free. We confirm strong emission lines in MATLAS-585 indicating ongoing star formation.
We find no clear evidence for emission in MATLAS-2094 in our spectra. However, for MATLAS-2103 we find [OII]3727, and weak H$\beta$ and [OIII]4959, 5007 emission. These emission lines suggest some ongoing star formation (in addition to an underlying older population). The redder spectral coverage of MUSE may explain our inability to confirm the emission lines reported by \citet{Heesters2023} in MATLAS-2094, but it is odd that the MUSE spectra do not reveal emission for MATLAS-2103. \cite{Heesters2023} found all three galaxies to be very old and of low metallicity, suggesting that their results refer to the underlying old stellar population and not the young starburst in each case. Of the three galaxies, our spectra allowed for reliable stellar population estimates only for MATLAS-585 and 2094.

Due to its ongoing star formation, we derive only a light-weighted stellar population for MATLAS-585. However, our stellar population bootstrapping produced a posterior probability distribution skewed toward the youngest allowable ages and lowest metallicities (see Appendix \ref{appendix:cornerfigs}, which may suggest that the errors are underestimated and that the best-fit light-weighted values may lie beyond the limits of the model grid. \citet{Heesters2023} reported a mean mass-weighted age and metallicity of 9.0$^{+2.9}_{-3.1}$ Gyr and --1.88$^{+0.17}_{-0.13}$ dex. MATLAS-585 is also part of the \cite{Buzzo2024} SED fitting study. They found an age of 8.9$^{+3.4}_{-4.3}$ Gyr and [M/H] = --1.36$^{+0.27}_{-0.16}$ dex.

For MATLAS-2094, we find  a mass-weighted age of 8.61$^{+2.25}_{-2.36}$ Gyr, and [M/H] of --0.80$^{+0.36}_{-0.35}$ dex. This age is in good agreement with \cite{Heesters2023} who report 8.2$^{+4.2}_{-0.8}$ Gyr, however, they find a much lower metallicity of --1.83$^{+0.14}_{-0.16}$ dex.

\citet{Heesters2023} report the stellar population of MATLAS-2103 to have an age of 11.3$^{+2.2}_{-2.5}$ Gyr and a metallicity of --1.69$^{+0.01}_{-0.22}$ dex.

\subsubsection{MATLAS-607} 

Here we report a new recessional velocity for MATLAS-607 confirming its association with PGC028887. Its spectrum reveals a strong [OII]3727 emission line but weak [OIII]5007 suggesting low ionisation (e.g. from shocks). We quote a light-weighted age of 1.9 Gyr and low metallicity of [M/H] = --1.2 dex.

\subsubsection{MATLAS-631}

MATLAS-631 is a blue galaxy that reveals an emission line dominated spectrum, with strong emission lines of [OII]3727, [OIII]4959, 5007 and H$\beta$, indicating likely ongoing star formation. The recession velocity derived from its absorption (and emission) lines places the galaxy at a distance of $\sim$145 Mpc, which is similar to the nearby galaxy PGC1422425.  \cite{Poulain2021} assigned a distance of 40.9 Mpc assuming an association with PGC029321 but did not quote a value for the angular size or surface brightness of MATLAS-631. We quote a light-weighted age of 0.36 Gyr and moderate metallicity of --0.47 dex.

\subsubsection{MATLAS-646}

We are able to derive both a new recessional velocity and stellar population parameters for MATLAS-646. We confirm association with the NGC 3156 group and derive mass-weighted age of 8.6 Gyr and metallicity [M/H] of --1.1 dex. 
This galaxy has clear tidal tails, as also noted by \citet{Paudel2023} from an apparent interaction with one of the nearby giant galaxies NGC~3166 or NGC~3169.
The dwarf also has a bright nucleus that is expected to become a free-floating ultracompact dwarf after its host galaxy is completely tidally disrupted (see \citealt{Jennings2015,Wang2023}).

\subsubsection{MATLAS-739}

We derive a recessional velocity for MATLAS-739 of 937$\pm20$ km/s which is in reasonable agreement with the mean velocity of the Leo Group, i.e. 969 km/s.  \cite{Poulain2021} associated MATLAS-739 with NGC 3377 of the Leo Group at a distance of 10.9 Mpc. 
The dwarf also has HST observations in \citet{Cohen2018}, who called it M96-DF11. In this study, they derived two distance estimates for the dwarf, both in reasonable agreement with each other; 9.5 $\pm$ 0.3 Mpc using the uniform luminosity of the brightest red giant stars (the `tip of the red giant branch' method), and 10.0 $\pm$ 1.3 Mpc using the variation of brightness caused by individual stars in the galaxy (the `surface brightness fluctuation' method).

\subsubsection{MATLAS-951}

MATLAS-951 is included in the SED study of \cite{Buzzo2024}. They find an old age of 9.2 Gyr and metallicity of [M/H] = --1.33. We are unable to derive stellar population parameters. Our recessional velocity suggests an association with the NGC 3640 group.
The dwarf also has an apparent nucleus and S-shaped tidal features that are likely due to an interaction with NGC~3640 \citep{Marleau2021}.

\subsubsection{MATLAS-1494}

The SED fitting study of \cite{Buzzo2024} assumed that MATLAS-1494 was associated with NGC 4690 \citep{Poulain2021} at a distance of 40.2 Mpc. 
Here we measure a recessional velocity of 1315 km/s suggesting a closer distance of $\sim$18 Mpc (for H$_0$ = 72 km/s/Mpc), and placing the galaxy in the field. This closer distance results in an effective radius of 1.16 kpc based on the angular size of 13.09 arcsec from \cite{Poulain2021}. Given the new size measurement, the galaxy should be reclassified from a UDG candidate to a (slightly smaller) dwarf galaxy. 

We are unable to derive reliable stellar population parameters from our KCWI spectrum. However, we conduct an SED fit using the photometry from  \cite{Buzzo2024} and following their method but now using the new distance of 18 Mpc. We derive a lower mass, slightly younger and more metal-rich galaxy with the closer distance, i.e. 
log M$_{\ast}$ = $7.33^{+0.15}_{-0.11}$ M$_{\odot}$,  [M/H] = $-0.99^{+0.46}_{-0.43}$ dex, 
age = $8.11^{+3.12}_{-2.34}$ Gyr.  
We report this updated stellar mass in Table 2. Although slightly smaller in half-light radius than a UDG, MATLAS-1494 represents an interesting case of an old passive dwarf galaxy than is relatively isolated. This would suggest that it is not quenched by any tidal interaction process. 

\subsubsection{MATLAS-1794}

Although we are unable to derive stellar population parameters for MATLAS-1794, we report a new recessional velocity of 1839 km/s which confirms its association with host galaxy NGC 5507 at a distance of 29 Mpc. We note that \cite{Marleau2024} find MATLAS-1794 to host fewer than one GC.

\subsubsection{MATLAS-1938}

Of the MATLAS galaxies in our sample, MATLAS-1938 has the richest estimated GC system of 28.5$\pm$5.7 \citep{Marleau2024}. Thanks to its bright nucleus, the galaxy has a velocity measured from SDSS of 1290 km/s, fully consistent with our own.  
Although \cite{Poulain2021} assumed it to be associated with NGC 5813 at a distance of 31.3 Mpc, 
this was revised by \cite{Marleau2024} who quote a distance of 17.8 Mpc. 
The Sb galaxy NGC 5806 (distance $\sim$ 25 Mpc) is both close on the sky and in velocity (1347 km/s) to MATLAS-1938. If we assume that MATLAS-1938 lies in the NGC 5846 group, along with NGC 5806, then the distance is $\sim$ 25 Mpc \citep{2001ApJ...546..681T}. 
At 25 Mpc the revised effective radius is 1.57 kpc and thus MATLAS-1938 meets the 
UDG definition. The closer distance used by \cite{Marleau2024} will result in a brighter assumed apparent magnitude of the universal GC luminosity function peak, than it would be at further distances, causing an underestimate of the GC system numbers. Due to the small difference in distance, the underestimation would likely not be significant.

The spectrum of MATLAS-1938 reveals weaker Balmer lines than most other galaxies in our sample. Indeed, we find an old mass-weighted age of 10.47$^{+1.38}_{-1.96}$ Gyr and [M/H] = --0.72$^{+0.15}_{-0.13}$ dex. 
\cite{Buzzo2025} found a very old age of 12.6$^{+0.20}_{-0.22}$ Gyr and [M/H] = --1.24$^{+0.01}_{-0.01}$ dex, using the \cite{Poulain2021} distance of 31.3 Mpc.

\subsubsection{MATLAS-2019}

Although not part of this study, MATLAS-2019 (also known as NGC5846\_UDG1), has been observed with KCWI \citep{Haacke2025}. They measured a galaxy velocity of 2150.7 km/s and a mean GC system velocity of 2153.7 km/s. This recessional velocity places MATLAS-2019 within the NGC 5846 group (see e.g., \citealp{Haacke2025}, Figure 5 for these recessional velocities in the context of the NGC~5846 group) at a distance of 26.5 Mpc. 
From deep HST imaging \citet{Danieli2022} resolved GC candidates to $\sim$0.7 mag below the GC luminosity function peak and used a colour criterion (without a size cut) to select the fainter GC candidates. From this they 
derived a remarkably rich GC system with 54$\pm$9 members. Of these 54 GC candidates, 20 are now spectroscopically-confirmed members \citep{Haacke2025}. We note that the shallower imaging of \cite{Marleau2024} found 38$\pm$7 GCs for an assumed distance of 20.3 Mpc.

This galaxy has measured ages and metallicities from \citet{Muller2020} and \citet{Heesters2023} which both appear to use data from the ESO program  0103.B-0635 (PI: Marleau). \citet{Muller2020} derived an age of 11.2$^{+1.8}_{-0.8}$ Gyr and a metallicity of [Z/H] = --1.33$^{+0.19}_{-0.01}$ dex. \citet{Heesters2023} derived an age of 13.5$^{+0.5}_{-0.2}$ Gyr and a metallicity of [Z/H] = --1.88$^{+0.13}_{-0.06}$ dex. i.e., \citet{Muller2020} derived stellar populations that are both younger and more metal-rich than \citet{Heesters2023}. It is worth noting that \citet{Heesters2023} employed a different data reduction and masking procedure than \citet{Muller2020} to ensure a consistent methodology across the 56 dwarfs analysed in their study.

\begin{figure}
\includegraphics[width = 0.55 \textwidth]{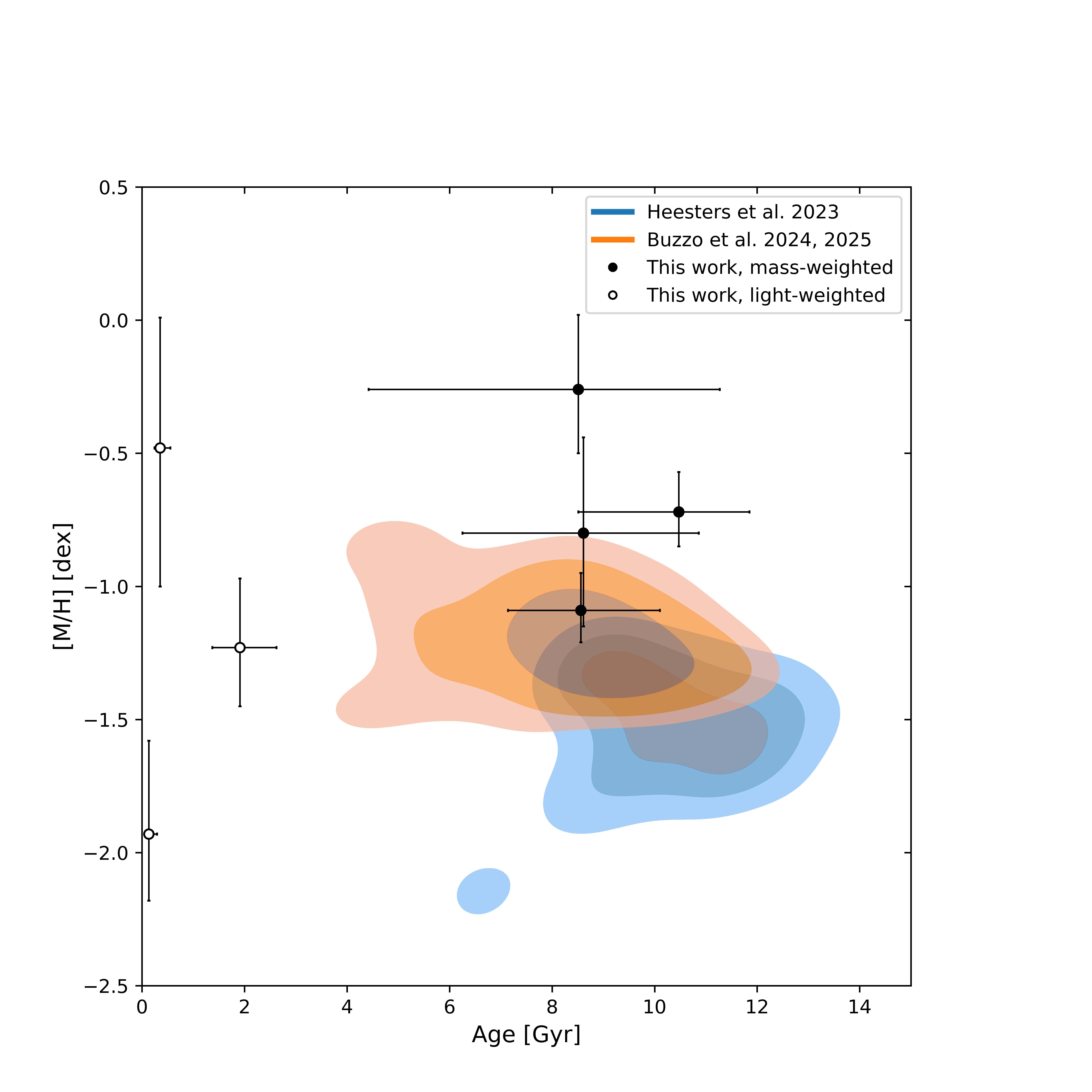}
\caption{Ages and metallicities ([M/H]) for MATLAS galaxies. Mass-weighted results presented in this work are shown with a filled black circle, and light-weighted results by an open circle.
Results from this work are compared to the distribution of ages and metallicities of MATLAS UDGs and dwarfs from the spectroscopic study of \cite{Heesters2023} (blue cloud) and the photometric study of \cite{Buzzo2024} (orange cloud). Our MATLAS galaxies have similar mass-weighted ages, on average, to the Buzzo et al. studies and are systematically more metal-rich than previous studies.}
\label{fig:amr}
\end{figure}

\subsection{Ages and Metallicities}

In Figure \ref{fig:amr} we show our derived spectroscopic ages and metallicities relative to the findings of \cite{Heesters2023} plus \cite{Buzzo2024} and \cite{Buzzo2025} from their large samples of MATLAS dwarf galaxies (i.e. UDGs and other dwarf galaxies). The \cite{Heesters2023} study is a spectroscopic one based on MUSE spectra, whereas the Buzzo et al. studies are based on photometry using SED fitting. On average, our galaxies have similar mass-weighted ages but are systematically more metal-rich than the Buzzo et al. studies. This offset had been previously identified in \citet{FerreMateu2023}, suggesting a bias toward lower metallicities from SED fitting. 

In comparison to the \cite{Heesters2023} sample, our sample, sometimes with large uncertainties, tends to be younger and more metal-rich. We note such a systematic effect for the two spectroscopic studies of MATLAS-2019 above. We also note the conclusions of \citet{Heesters2023}, which speculated that their reliance on the Calcium Triplet for metallicity measurements in their work may have been insufficient for a robust metallicity measurement. Our results of higher metallicities may help alleviate their offset found for MATLAS dwarfs from the universal mass-metallicity relationship of \citet{Kirby2013}. Larger samples will better define the distribution of stellar population parameters for UDGs and classical dwarfs in low-density environments.

\begin{table*}
\caption{Summary of Results.}\label{literature-data-table}
{\tablefont\begin{tabular}{@{\extracolsep{\fill}}lcrrrrrrrrr}
\toprule
Target ID &  Host & D$_{host}$ [Mpc] & V$_{host}$ [km/s] & V$_{galaxy}$ [km/s] & Age [Gyr] & [M/H] [dex] & N$_{GC}$ & log~M$_{\ast}$ [M$_{\odot}$] \\

\hline

MATLAS-585 & IC~0560 &  27.0 & 1853 & 1902 $\pm$ 13 & 0.13$^{+0.16}_{-0.07}$ & -1.93$^{+0.35}_{-0.25}$ &  10.50 $\pm$ 4.12 & 7.53\\

MATLAS-607  & PGC028887 &  41.0 & 2833 & 2976 $\pm$ 10 & 1.91$^{+0.71}_{-0.54}$ & -1.23$^{+0.26}_{-0.22}$  & -- & --\\

MATLAS-631  & {\it PGC1422425} &  145 & 10470 & 10266 $\pm$ 59 & 0.35$^{+0.20}_{-0.11}$ & -0.48$^{+0.48}_{-0.52}$ &  -- & --\\

MATLAS-646  & NGC~3156 &  21.8 & 1338 & 1418 $\pm$ 7 & 8.56$^{+1.54}_{-1.42}$ & -1.09$^{+0.14}_{-0.12}$ & -- & --\\

MATLAS-739  & NGC~3377  &  10.9 & 690 & 937 $\pm$ 20 & -- & -- & --& --\\

MATLAS-951 & NGC~3640 &  26.3 & 1298 & 1695 $\pm$ 23 & -- & -- & -- & 8.25\\

MATLAS-1494 &  {\it Isolated?} & --  & -- & 1315 $\pm$ 22 & -- & -- & -- & 7.33\\

MATLAS-1794 & NGC~5507  & 29.0 & 1851 & 1839 $\pm$ 12 &  -- & -- &  0.46 $\pm$ 2.82 & 7.40 \\

MATLAS-1938 &  {\it NGC~5806} & 25.0 & 1347 & 1303 $\pm$ 8 &  10.47$^{+1.38}_{-1.98}$ & -0.72$^{+0.15}_{-0.13}$ & 28.5 $\pm$ 5.71 & --\\

MATLAS-1957 &  NGC~5813  & 31.3 & 1956 & 2371 $\pm$ 17 & 8.51$^{+2.78}_{-4.09}$ & -0.26$^{+0.28}_{-0.24}$ & -- & 8.00\\

MATLAS-2094 &  NGC~6010 &  30.6 & 2381 & 1988 $\pm$ 15 & 8.61$^{+2.25}_{-2.36}$ & -0.80$^{+0.36}_{-0.35}$ & -- & --\\

MATLAS-2103 &  NGC~6017 &  35.8 & 1788 & 1713 $\pm$ 16 & -- & -- & -- & 8.21
\botrule
\end{tabular}}
\begin{tabnote}
Notes: MATLAS galaxy ID (column 1), associated host galaxy with revised hosts in italics (column 2), distances and recession velocity to the host galaxy as adopted from the literature (columns 3,4), radial velocity of the galaxy as measured in this work (column 5), measured age and metallicity in this work (columns 6,7) with light-weighted values for MATLAS-585, 607 and 631 and mass-weighted for all others, globular cluster numbers (column 8) from \cite{Marleau2024} and stellar masses from \cite{Buzzo2025}.

\end{tabnote}

\end{table*}

\section{Conclusions}

Of the 56 MATLAS dwarfs with recession velocities measured by \cite{Heesters2023} and a further 3 from \cite{Buzzo2024}, here we add an additional 12 (although several are in common with these literature studies). 
The relative fraction of MATLAS galaxies associated with their previously proposed host galaxy is around 80\%.
Here we find one UDG candidate, MATLAS-1494, to be located at a closer distance than the suggested host galaxy (NGC~4690). We now place it at 1315 km/s, or a distance of $\sim$18 Mpc, and in a relatively isolated environment. Given its angular effective radius of 13.09 arcsec \citep{Poulain2021}, its physical effective radius is calculated to be 1.16 kpc and so no longer a UDG candidate but is now a more regular dwarf galaxy. New SED fitting indicates it is old and passive. This is a challenge to theories that invoke tidal interactions in order to quench dwarf and/or UDG galaxies. 

\cite{Heesters2023} found 17/56 (=30\%) to reveal emission lines, and we find a similar fraction with 3 (strong) and a further 1 (weak) out of 12. Further work is needed to understand whether any of the emission lines are indicative of an AGN (a low fraction is expected given the typical stellar mass of 10$^8$ M$_{\odot}$). 

We also report mean ages and metallicities for 7 MATLAS dwarf galaxies. Compared to previously reported stellar population studies of \cite{Heesters2023}, \cite{Buzzo2024} and \cite{Buzzo2025},   
we generally find younger ages and higher metallicities. The Buzzo et al. studies were based on global photometry and a spectral energy distribution fit with the (reasonable) assumption that the MATLAS galaxies were at the same distance as their assumed host. The \cite{Heesters2023} work was based on fitting a spectrum from the MUSE instrument which excludes key absorption lines blueward of H$\beta$ and is mainly reliant on the CaT lines. Our results of higher metallicities has the potential to reduce the offset of MATLAS dwarf galaxies from the dwarf galaxy stellar mass--metallicity relationship found by \citet{Heesters2023}.
More work (e.g. detailed star formation histories), and larger samples, are required to determine whether or not these results extend to the wider MATLAS dwarf population. Nevertheless, our small sample suggests that UDGs and dwarf galaxies in low density environments are a `mixed bag', ranging from some with current star formation, to post starbursts with star formation a few Gyr ago, to old fully-quenched galaxies (as seen for UDGs in clusters).

\section*{Data Availability}
The KCWI data presented are available via the Keck Observatory Archive (KOA): \url{https://www2.keck.hawaii.edu/koa/public/koa.php} 18 months after observations are taken.


\section{Acknowledgements}

We thank the referee for several useful comments. 
We thank L. Haacke for help with the observations. 
DAF and JPB thank the ARC for financial support via DP220101863.
AJR was supported by National Science Foundation grant AST-2308390.
Here we make use of NED (https://ned.ipac.caltech.edu/) and
Qfitsview (https://www.mpe.mpg.de/ott/QFitsView/).
The data presented herein were obtained at Keck Observatory, which is
a private 501(c)3 non-profit organization operated as a scientific partnership among the California Institute of Technology, the University of California, and the National Aeronautics and Space Administration. The Observatory was made possible by the generous financial support of the W. M. Keck Foundation. The authors wish to recognize and acknowledge the very significant cultural role and reverence that the summit of Maunakea has always had
within the Native Hawaiian community. We are most fortunate to have the opportunity to conduct observations from this mountain. This research made use of Montage. It is funded by the National Science Foundation under Grant Number ACI-1440620, and was previously funded by the National Aeronautics and Space Administration's Earth Science Technology Office, Computation Technologies Project, under Cooperative Agreement Number NCC5-626 between NASA and the California Institute of Technology.
Parts of this work were performed on the OzSTAR national facility at Swinburne University of Technology. The OzSTAR program receives funding in part from the Astronomy National Collaborative Research Infrastructure Strategy (NCRIS) allocation provided by the Australian Government, and from the Victorian Higher Education State Investment Fund (VHESIF) provided by the Victorian Government.



\newpage
\appendix
\renewcommand{\thefigure}{\thesection.\arabic{figure}}
\onecolumn

\section{Galaxy Masks}\label{appendix:galaxymasks}

\begin{figure*}[ht]
    \centering
    \begin{subfigure}[b]{0.8\textwidth}
         \includegraphics[scale=0.5]{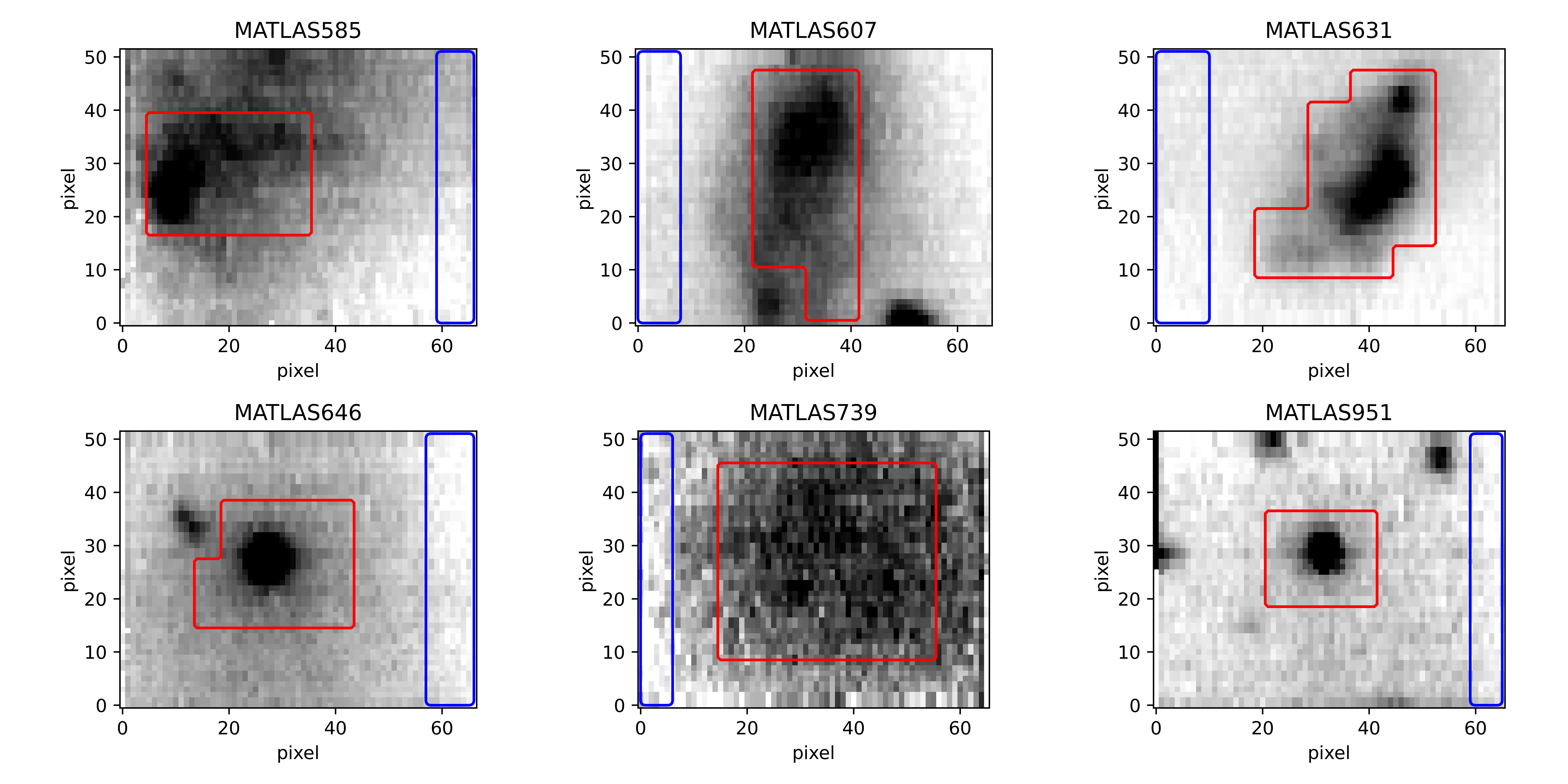}
    \end{subfigure}
    
    \begin{subfigure}[b]{0.8\textwidth}
         \includegraphics[scale=0.5]{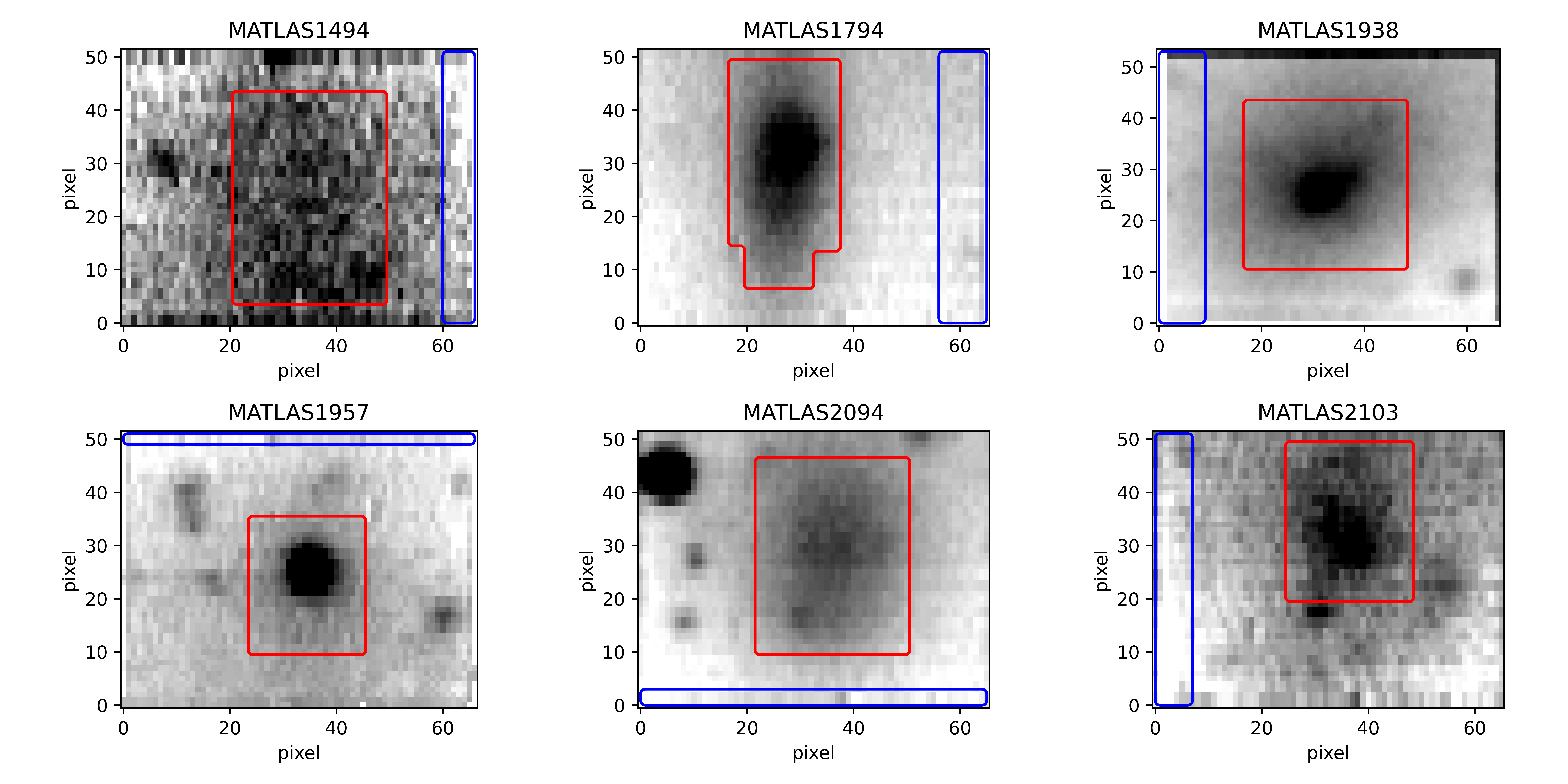}
    \end{subfigure}

    \caption{Cutouts of the 12 MATLAS galaxies studied in this work. They are shown in inverted grayscale, scaled between the 5th and 98th image percentiles to enhance the contrast. The red box outlines the on-source region identified as the source, while the blue box indicates the {\color{blue}{on-sky}} region used for background estimation. A 1D spectrum for each galaxy was computed in QFitsView, which performed background subtraction by subtracting the average flux in the on-sky region from the on-source region.}

\end{figure*}

\newpage

\section{Corner Plots from Stellar Population Bootstrapping}\label{appendix:cornerfigs}
\begin{figure*}[ht]
    \centering
    \begin{subfigure}[b]{0.48\textwidth}
        \centering
        \includegraphics[scale=0.6]{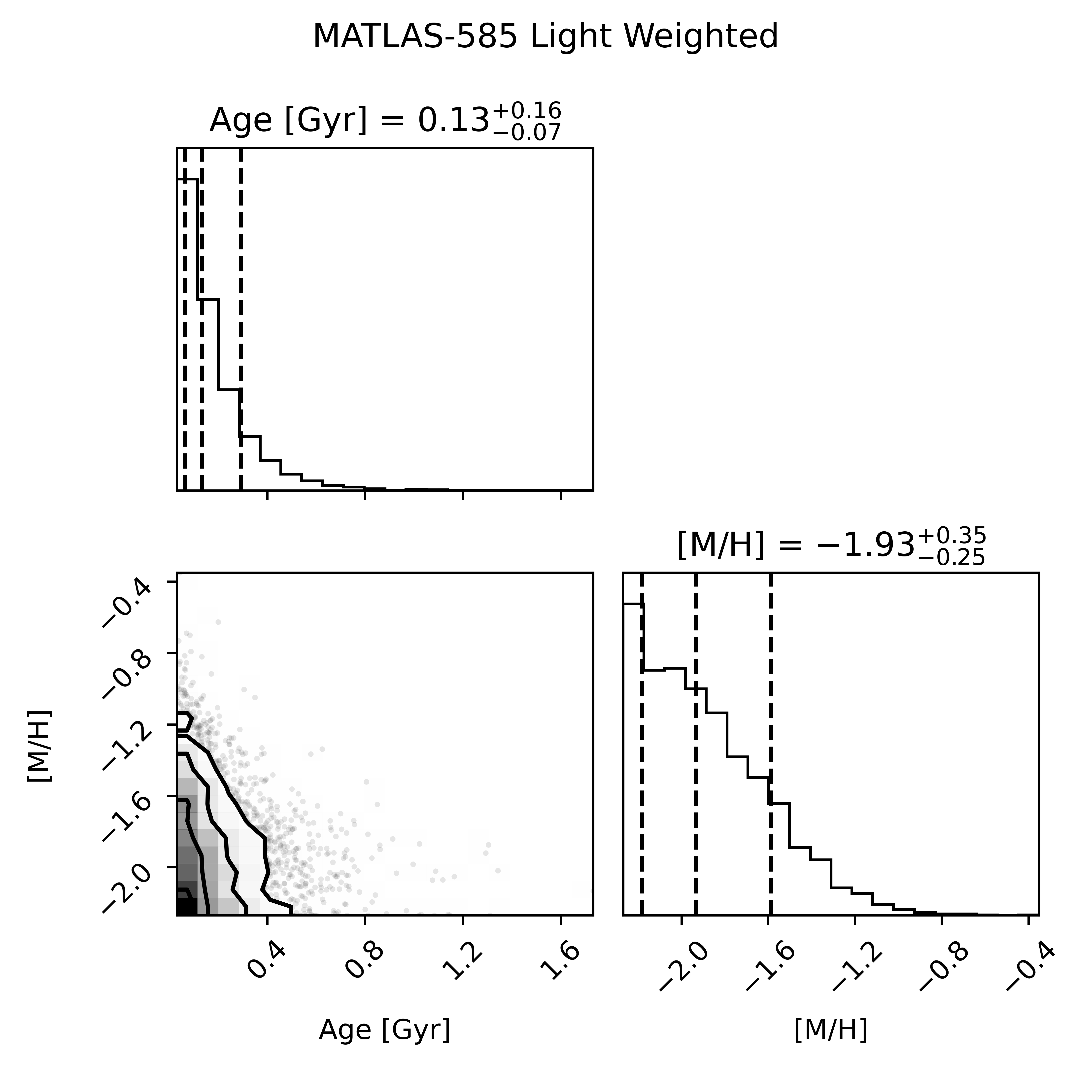}
        \caption{}
    \end{subfigure}
    \begin{subfigure}[b]{0.48\textwidth}
        \centering
        \includegraphics[scale=0.6]{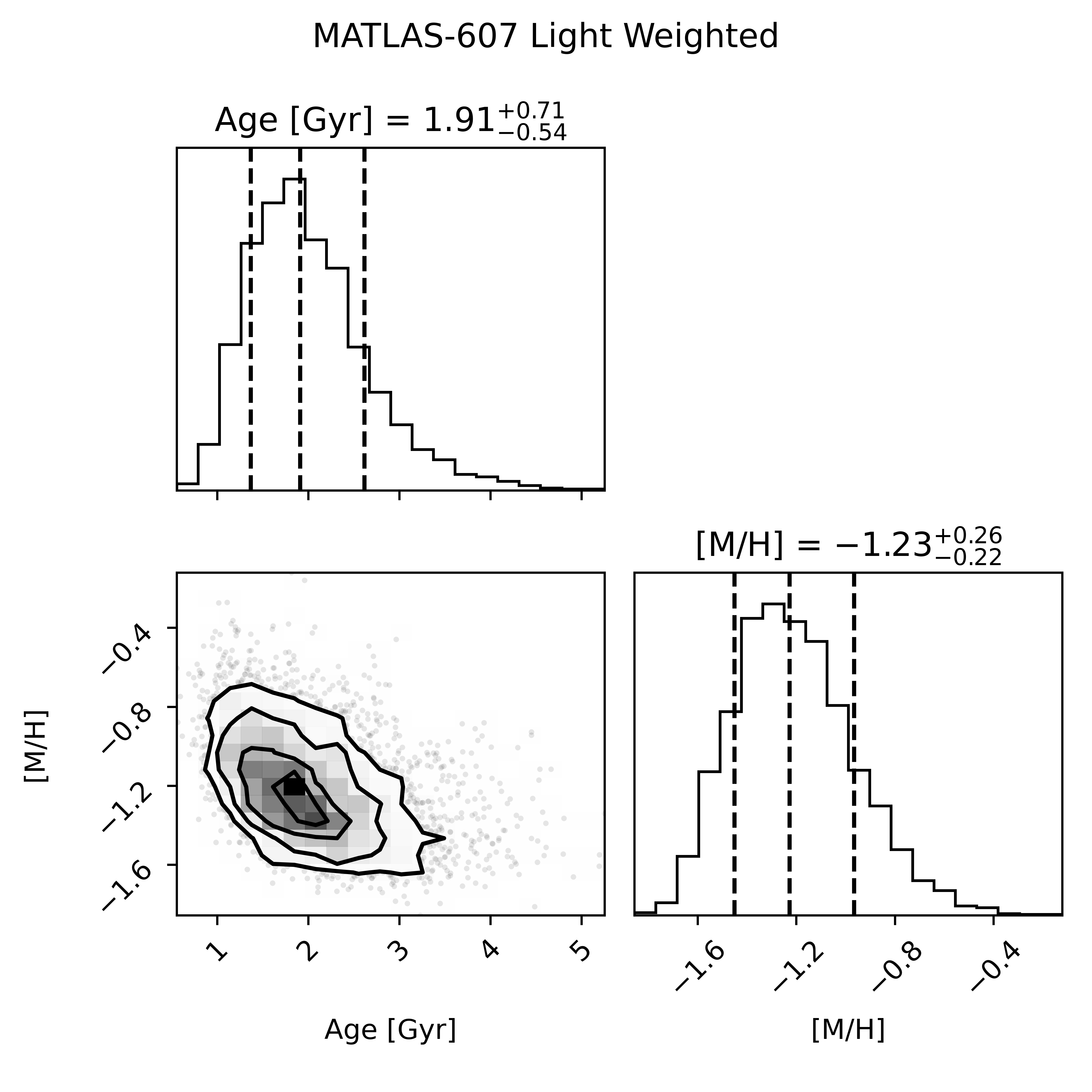}
        \caption{}
    \end{subfigure}
    
    \vspace{0.5cm}
    
    \begin{subfigure}[b]{0.48\textwidth}
        \centering
        \includegraphics[scale=0.6]{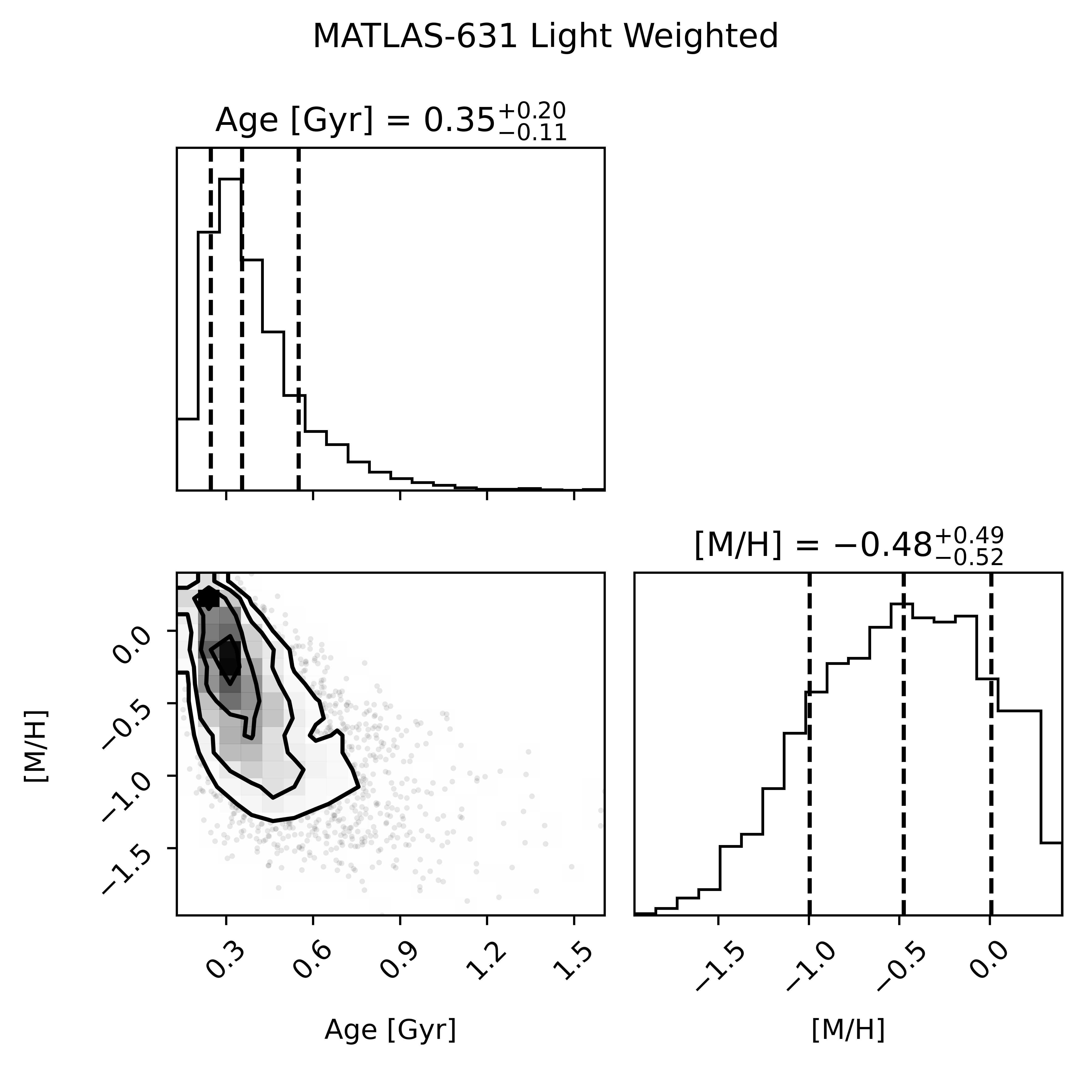}
        \caption{}
    \end{subfigure}
    \begin{subfigure}[b]{0.48\textwidth}
        \centering
        \includegraphics[scale=0.6]{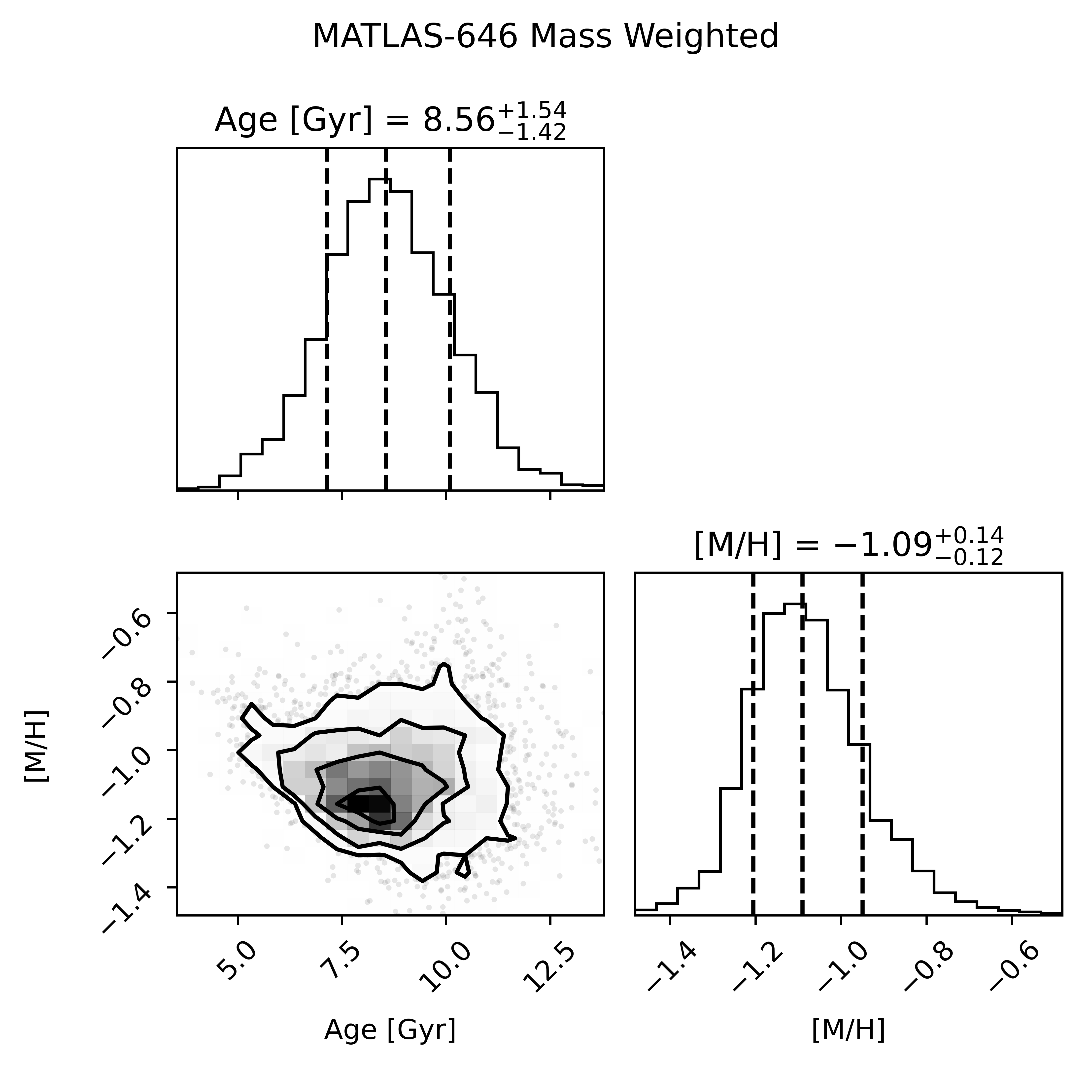}
        \caption{}
    \end{subfigure}

    \caption{Corner plots of the results of 10000 bootstrap realisations of the stellar population fitting for MATLAS-585, 607, 631, and 646. In each figure, the diagonal panels show the distributions of ages and metallicities as histograms, while the off diagonal panel shows a contour plot of their correlation. Results are quoted as the 16th, 50th, and 84th quantiles, which are marked with vertical dashed lines in the diagonal panels. The results for MATLAS-646 shown in (d) are mass-weighted, while the results for the other three are light-weighted.}

\end{figure*}
    
\begin{figure*}[ht]
    \centering
    \begin{subfigure}[b]{0.48\textwidth}
        \centering
        \includegraphics[scale=0.6]{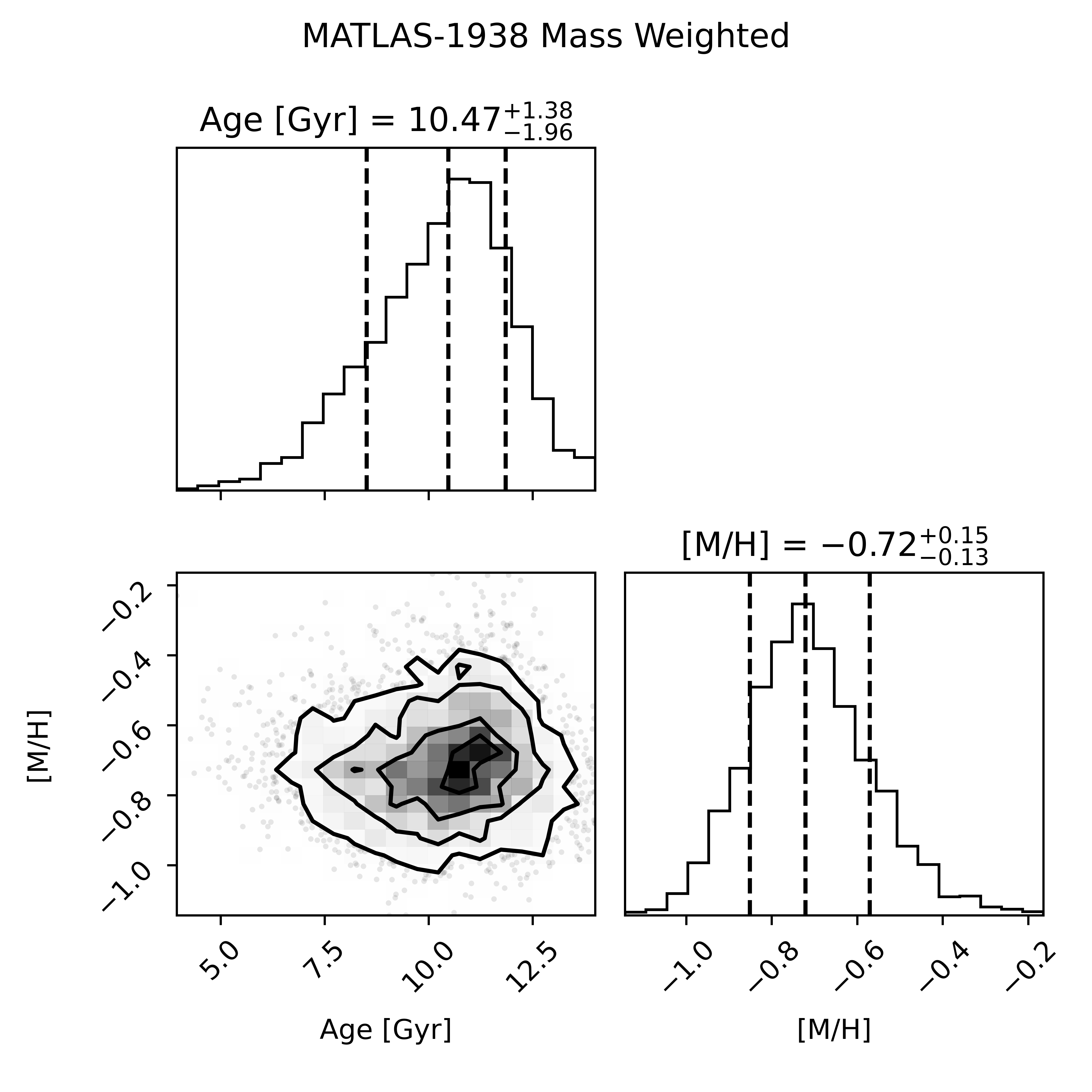}
        \caption{}
    \end{subfigure}
    \begin{subfigure}[b]{0.48\textwidth}
        \centering
        \includegraphics[scale=0.62]{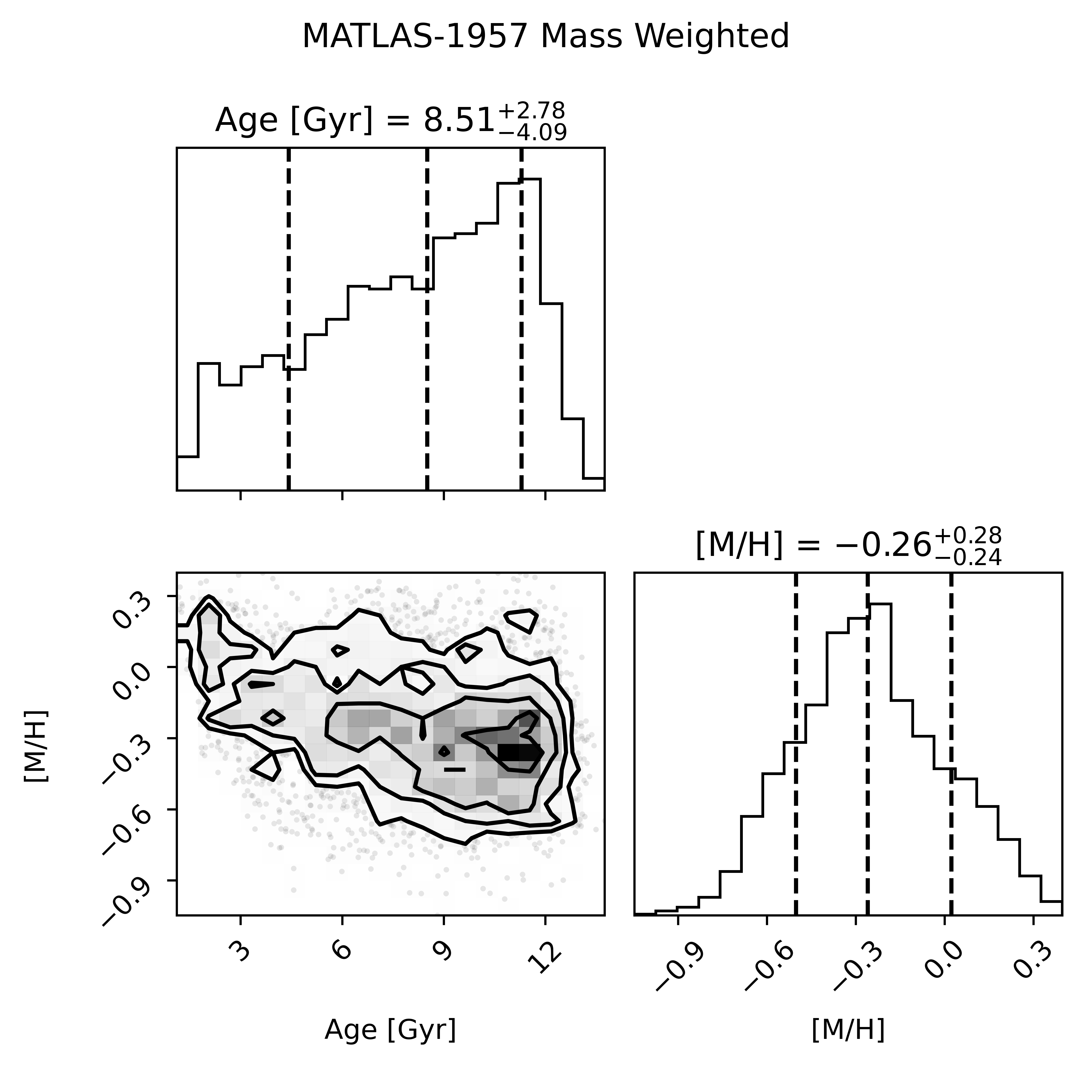}
        \caption{}
    \end{subfigure}
    
    \vspace{0.5cm}
    \begin{subfigure}[b]{0.48\textwidth}
        \centering
        \includegraphics[scale=0.62]{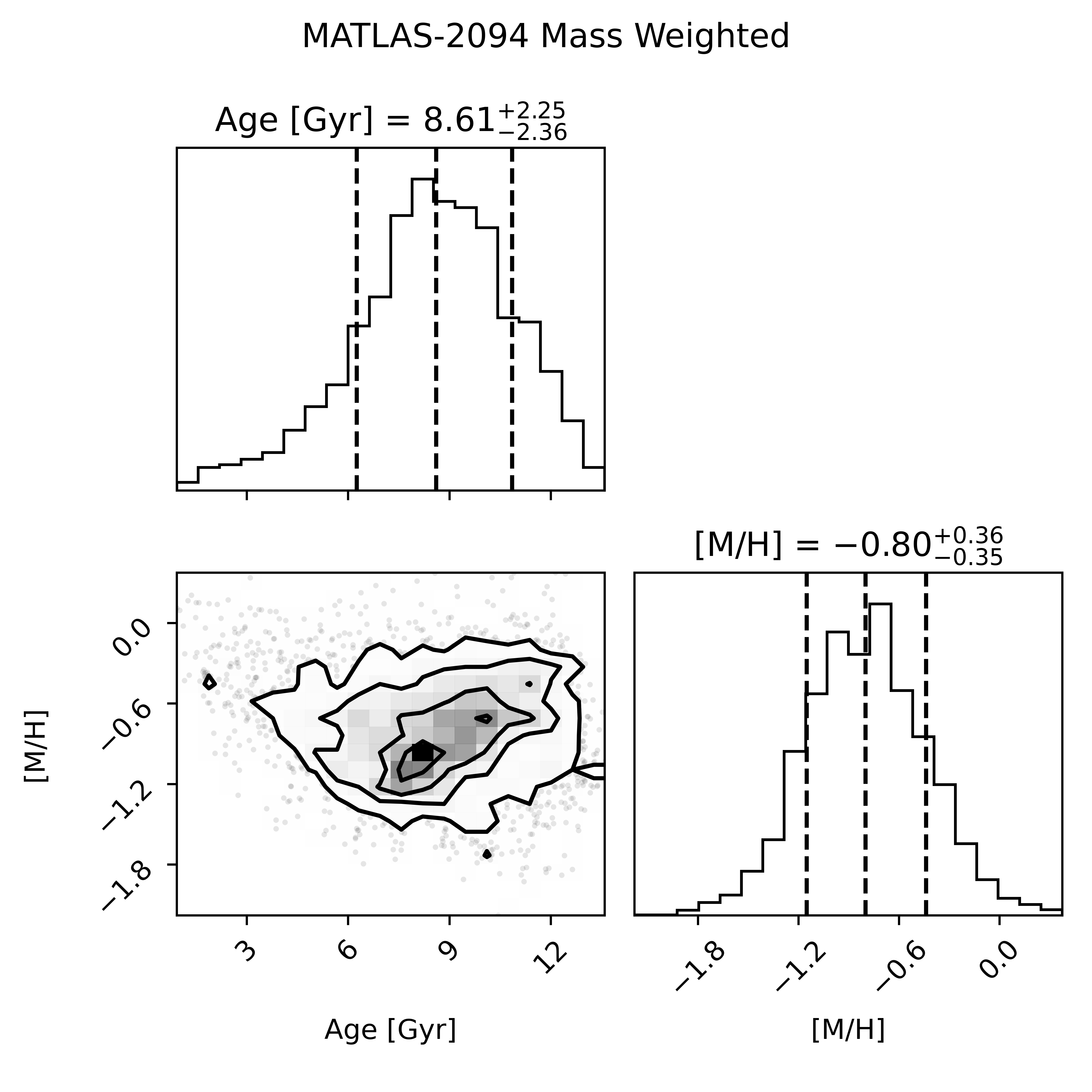}
        \caption{}
    \end{subfigure}
    
    \caption{Corner plots of the results of 10000 bootstrap realisations of the stellar population fitting for MATLAS-1938, 1957, and 2094. In each figure, the diagonal panels show the distributions of ages and metallicities as histograms, while the off-diagonal panel shows a contour plot of their correlation. Results are quoted as the 16th, 50th, and 84th quantiles, which are marked with vertical dashed lines in the diagonal panels. All results are mass-weighted.}
\end{figure*}

\end{document}